\documentclass{aa}    
\usepackage{graphicx}
\def\ut#1{\mathop{\vtop{\ialign{##\crcr
     $\hfil\displaystyle{#1}\hfil$\crcr\noalign
     {\kern1pt\nointerlineskip}\hbox{$\hfil\sim\hfil$}\crcr
     \noalign{\kern1pt}}}}}

\def\undersymbol#1#2{\mathop{\vtop{\ialign{##\crcr
     $\hfil\displaystyle{#2}\hfil$\crcr\noalign
     {\kern1pt\nointerlineskip}\hbox{$\hfil#1\hfil$}\crcr
     \noalign{\kern1pt}}}}}

\begin{document}
\thesaurus{11.02.2, 02.02.01}
\title{Astrophysical implications of binary black holes
in BL Lacertae objects}
\author{
    F. De Paolis
\inst{},
    G. Ingrosso
\inst{} and
  A. A. Nucita
\inst{}
     }
\institute{Dipartimento di Fisica, Universit\`a di Lecce and INFN
Sezione di Lecce, Via Arnesano CP 193 I-73100 Lecce, Italy }
\offprints{F. De Paolis}
\date{Received 2 July 2001 / Accepted  25 March 2002}
\authorrunning{F. De Paolis et al.}
\titlerunning{Astrophysical implications of binary
black holes} \maketitle
\begin{abstract}

Some BL Lacertae objects show a periodic behaviour in their light
curves that is often attributed to the orbital motion of a central
binary black hole system. On this basis, and assuming a circular
orbit, Rieger and Mannheim (\cite{rm}) have recently proposed a
method to determine the orbital parameters of the binary system
from the observed quantities, {\it i.e.} the signal periodicity,
the flux ratio between maximum and minimum signal and the power
law spectral index of the photon flux.

However, since these binary black holes are expected to originate
from galactic mergers, they could well be on eccentric orbits,
which might not circularize for a substantial time. We therefore
generalize the treatment proposed by Rieger and Mannheim
(\cite{rm}) by taking into account the effect of the orbital
eccentricity of the binary system. We apply the model to three
well-observed Markarian objects: MKN 501, MKN 421 and MKN 766 that
most likely host a binary system in their centers. Some
astrophysical implications of this model are also investigated
with particular emphasis to the gravitational radiation emission
from the binary black holes. Under particular conditions (e.g.
for some values of the orbital separation, eccentricity and
Lorentz factor) one can obtain signals above the sensitivity
threshold of the LISA detector. 
\end{abstract}

\section{Introduction}
BL Lacertae objects, also known as Markarian objects (hereafter
MKNs), belong to the class of active galaxies. According to the
well-established unified model on radio-loud active galactic
nuclei (see Urry and Padovani \cite{up}), these objects are
thought to be dominated by relativistic jets seen at small angles
to the line of sight. The structure of the relativistic jets is
largely unknown and the typical smooth and fairly featureless
spectra of such objects may be reproduced by models making use of
very different assumptions. Combining spectral and temporal
informations greatly constrains the available models.

Up to now, several astrophysical phenomena have been attributed
to binary black holes, like precession of jets (Begelman et al.
\cite{bbr}), misalignment (Conway and Wrobel \cite{cw}), periodic
outburst activity in the quasar OJ $288$
(Sillinp$\ddot{a}\ddot{a}$ et al. \cite{shv}, Letho and Valtonen
\cite{lv}) and precession of the accretion disk under
gravitational torque (Katz \cite{katz}).

It has been recently observed that some MKN objects show a
periodic behaviour in the radio, optical, X-ray and $\gamma$-ray
light curves that  is possibly related to the presence of a
massive binary black hole creating a jet either aligned along the
line of sight or interacting with an accretion disk (Yu
\cite{yu}).
Therefore, the search for X-ray/$\gamma$-ray
variability can be considered as a method to probe
the existence of a massive binary black hole in the center of
a galaxy.

Massive black hole systems are expected to be fairly common in
the Universe as a result of merging between galaxies. Indeed, in the
framework of the unified model for the morphological evolution of
galaxies (see {\it e.g.} White \cite{white}) giant ellipticals,
which are the host galaxies of most MKN objects, appear
to be the result of mergers between spirals. On the other hand,
it is now well established that galaxies generally contain
massive black holes in their nuclei (Rees \cite{rees}, Kormendy
and Richstone \cite{kr}, Richstone et al. \cite{rab}). Therefore,
merging would naturally lead to the formation of massive binary
black holes (Begelman et al. \cite{bbr}).

At least three MKN objects ({\it i.e}  MKN 501, MKN 421 and MKN
766) are particularly well studied at high energies, revealing a
possible periodic behaviour in their light curves.

MKN 501, at $z=0.034$, shows a clear well-correlated 23 day
periodicity in $X$-ray and TeV energy bands with an observed TeV
flux ratio $f\simeq$ 8 between the maximum and minimum of the
signal (Protheroe et al. \cite{pbf}, Hayashida et al. \cite{hhi},
Kranich et al. \cite{kdk}, Nishikawa et al. \cite{nhc}), while
evidence for correlations in the optical U-band is rather weak
(Catanese et al. \cite{cbb} , Aharonian  et al. \cite{aab}). It
has also been suggested that the complex morphology of the jet
and the peculiar behaviour of its spectral energy distribution
are probably related to the presence of a massive binary black
hole (Conway and Wrobel \cite{cw}, Villata and Raiteri \cite{vr}).

MKN 421, at $z=0.031$, is the brightest BL Lacertae object at
$X$-ray and UV wavelengths and it is the first extragalactic
source discovered at TeV energies (Punch et al. \cite{pac}). This
nearby source, which has been recently observed by the XMM-Newton
(see Brinkmann et al. \cite{bsg}) and by BeppoSax (Maraschi et al.
\cite{mft}) satellites, shows remarkable $X$-ray variability
correlated with strong activity at TeV energies (George et al.
\cite{gwb}) on a time-scale of $\simeq 10^{4}$ s (Maraschi et al.
\cite{mft}) and with a flux ratio $f \simeq 2$.

X-ray observations of the nearby MKN 766, at $z=0.013$, have been
performed by the XMM-Newton satellite (Boller, Keill,
Tr$\ddot{u}$mper et al. \cite{bkt}). These observations have
revealed the presence of a strong $X$-ray periodic signal with
frequency $\simeq 2.4\times 10^{-4}$ Hz and flux ratio $f\simeq
1.3$.

Based on the assumption that the periodic observed light curve of
MKN 501 is related to the presence of a binary system of black
holes (one of which emits a jet), Rieger and Mannheim (\cite{rm})
have proposed a method to determine the physical parameters of
the binary system from the observed quantities, i. e. signal
periodicity, flux ratio between maximum and minimum signal and
power law spectral index. They adopted the simplifying
assumptions that {\it i)} the two massive black holes move on
circular orbits around the common center of mass and {\it ii)}
the binary separation $a$ is such that the gas dynamical
time-scale $T_{gas}$ is equal to the time-scale for gravitational
radiation $T_{gw}$ (see also Begelman, Blandford and Rees
\cite{bbr}).

However, if binary black holes originated from galactic mergers,
they could be on eccentric orbits and eccentricity values up to
$0.8$-$0.9$ are not necessarily too extreme (Fitchett
\cite{fitchett}). Of course, due to gravitational wave emission,
orbits tend to circularize but this happens within a time-scale of
the same order of magnitude as the merging time-scale (Peters
\cite{peters}, Fitchett \cite{fitchett}). Therefore, if a massive
binary black hole is found at the center of a galaxy, it may
happen that the constituting black holes are still on eccentric
orbits, in which case the method proposed by Rieger and Mannheim
(\cite{rm}) does not hold.

As far as point {\it ii)} is concerned, we note that the
assumption that nowadays $T_{gas}$ is equal to $T_{gw}$ is
arbitrary. Indeed, it is possible that $T_{gas}=T_{gw}$ at some
time in the past, but since then the binary system evolution was
driven by gravitational wave emission which shrunk the orbital
separation faster (see Section 2).

The aim of the present paper is to generalize the model proposed
by Rieger and Mannheim (\cite{rm}) by taking into account the
effects of the orbital eccentricity  of the binary system and
considering the binary separation $a$ as a free model parameter.
This is done in Section 2, where the model is also applied to
three well-observed MKN objects (MKN 501, MKN 421 and MKN 766)
obtaining a set of possible values for the free model parameters
$\gamma_b$, $a$ and $e$ corresponding to different values for the
masses $m$ and $M$ of the two black holes. In Section $3$ we
study the gravitational waves emitted by the binary systems at
the center of the above-quoted MKN objects and analyze the
possibility of detecting them by the LISA gravitational wave
interferometer. Our conclusions are presented in Section 4.

\section{Binary black hole model for BL Lacertae objects}

In the commonly-accepted evolutionary scenario, giant elliptical
galaxies are believed to be the product of mergers between spiral
galaxies. Since each galaxy likely contains a central black hole
with mass 10$^6$-10$^9$ $M_{\odot}$ (Richstone, Ajhar and Bender
\cite{rab}), merging would sometimes produce binary black hole
systems separated by a typical distance of $0.1-1$ pc (Begelman,
Blandford and Rees \cite{bbr})).

We assume that binary black holes exist in the center of many MKN
objects and in particular in those of MKN 501, MKN 421 and MKN
766. As in Rieger and Mannheim (\cite{rm}), the periodicity in
the flaring state, observed towards these objects, is assumed to
be the consequence of the orbital motion of a relativistic jet in
the binary black hole. Therefore, the observed signal periodicity
has a geometrical origin, being a consequence of Doppler-shifted
modulation. Let us further assume that the jet is emitted by the
less massive black hole and that the nonthermal
$X-ray$/$\gamma$-ray radiation propagates outwards from the core
along the jet with Lorentz gamma factor $\gamma _b$. The observed
flux modulation due to Doppler boosting can be written as
\begin{equation}
S(\nu) = \delta ^{3+\alpha}S^{\prime}(\nu)~, \label{flux}
\end{equation}
where $\alpha$ is the source spectral index \footnote{ Values of
the power law index $\alpha$ for the three MKNs of interest are
found to be 1.2, 1.7 and 2.11 for MKN 501, MKN 421 and MKN 766,
respectively. For more details see Rieger and Mannheim
(\cite{rm}), Guainazzi, Vacanti, Maizia et al. (\cite{gvm}) and
Boller, keill, Tr$\ddot{u}$mper et al. (\cite{bkt}).} and the
Doppler factor is given by
\begin{equation}
\delta = \frac{\sqrt{1-(v_z^2+ v_{ls}^2)/c^2}}{1-(v_z \cos i +
v_{ls} \sin i )/c}~. \label{doppler}
\end{equation}
Here, $v_z$ is the outflow velocity in the direction of the total
angular momentum, $i$ is the inclination angle between the jet
axis and the line of sight and $v_{ls}$ is the component of the
less massive black hole velocity along the line of sight.

Let $M$ and $m$ represent the masses of the primary and the
secondary black hole, respectively. Conventionally, the problem
of predicting the motion of a system of two gravitationally
interacting point masses is simplified by considering the
equivalent system. This consists of a mass $M_t=M+m$ fixed in
space and acting on a reduced mass $\mu =Mm/(M+m)$ that orbits
around it. We assume that the reduced mass moves around the total
mass on an elliptic orbit with semi-major axis $a$ and
eccentricity $e$. Let $(r, \theta)$ be the polar coordinates of
$\mu$ relative to $M_t$, so that the equation of the orbit is
given by
\begin{equation}
r=\frac{a(1-e^2)}{1+e\cos \theta}~. \label{elliorbi}
\end{equation}
It is straightforward to demonstrate that the component of the
reduced mass velocity along the line of sight is given by (Smart
\cite{smart})
\begin{equation}
v_{\mu,ls} = \sqrt{\frac{G(M+m)}{a(1-e^2)}}\sin \theta~.
\label{vreducedmass}
\end{equation}
The component of the less massive black hole velocity along the
line of sight is thus given by
\begin{equation}
v_{ls} = \frac{M}{M+m}v_{\mu,ls}~,
\end{equation}
which can be rewritten as
\begin{equation}
v_{ls} = R_a \Omega _a \frac{\sin \theta}{(1-e^2)^{\frac{1}{2}}}~,
\label{vlessmassivebh}
\end{equation}
where $R_a=Ma/(M+m)$ and $\Omega_a =[G(M+m)/a^3]^{1/2}$ is the
Keplerian orbital frequency if the mass $m$ moves on a circular
orbit of radius $a$ around $M$. The velocity $v_{ls}$ is maximal
for $\theta = \pi /2$ and minimal for $\theta = 3\pi /2$,
corresponding through equation (\ref{doppler}) to the two possible
values $\delta_{max}$ and $\delta_{min}$ of the Doppler factor.

With the assumption that the periodicity in the observed signal
is due to the orbital motion of the binary black hole,
from equation (\ref{flux}) one obtains
the condition $\delta _{max}/ \delta_{min} \simeq
f^{1/(3+\alpha)}$, where $f$ is the observed maximum to minimum
flux ratio. Consequently, by using equations (\ref{doppler}) and
(\ref{vlessmassivebh}) we have
\begin{equation}
\Omega _a R_a =
\frac{f^{1/3+\alpha}-1}{f^{1/3+\alpha}+1}\left(\frac{1}{\sin i}
-\frac{v_z}{c}\cot i \right)(1-e^2)^{\frac{1}{2}}c~,
\label{omegar}
\end{equation}
which, in the limit $e\rightarrow 0$, reduces to equation ($4$) in
Rieger and Mannheim (\cite{rm}).

As a consequence of the binary orbital revolution around the
center of mass, the observed signal period $P_{obs}$ is related to
the keplerian period $P_k = 2\pi/\Omega_a$ by (Rieger
and Mannheim, \cite{rm})
\begin{equation}
P_{obs}= (1+z) \left(1- \frac{v_z}{c}\cos i \right)P_k~,
\label{observedperiodicity}
\end{equation}
where $i\simeq 1/\gamma _b$ and $v_z\simeq
c(1-1/\gamma_b^2)^{1/2}$ (Spada \cite{spada}).
From equations
(\ref{omegar}) and (\ref{observedperiodicity}), one obtains the
following relation
\begin{equation}
\begin{array}{l}
\displaystyle{\frac{M}{(m+M)^{2/3}}=
\frac{P_{obs}^{1/3}}{[2\pi(1+z)G]^{1/3}}\frac{c}{\sin i}}\times \\ \\
~~~~~~~~~~~~~~~~\displaystyle{
\frac{f^{1/3+\alpha}-1}{f^{1/3+\alpha}+1}\left(1
-\frac{v_z}{c}\cos i \right)^{2/3}} (1-e^2)^{1/2}~,
\end{array}
\label{massratio1}
\end{equation}
which, in the limit $e\rightarrow~0$, reduces to equation ($8$) of
Rieger and Mannheim (\cite{rm}).

Moreover, from equation (\ref{observedperiodicity}) one has
\begin{equation}
m+M =\left[\frac{2\pi(1+z)\left(1-\frac{v_z}{c}\cos
i\right)}{P_{obs}}\right]^2 \frac{a^3}{G}~, \label{massratio2}
\end{equation}
where $a$ represents the semi-major axis of the binary system.

Therefore, the allowed primary and secondary black hole masses $m$
and $M$ are obtained by solving simultaneously equations
(\ref{massratio1}) and (\ref{massratio2}). In these equations
$\gamma_b$, $e$ and $a$ have to be considered as free model
parameters to be determined by the observed data ($P_{obs}$, $f$
and $\alpha$) towards the three MKK of interest. In this way, we
can estimate the masses of the black holes supposed to be at the
centers of MKN 501, MKN 421 and MKN 766.

We mention that equation (\ref{massratio2}) is different from the
corresponding one in Rieger and Mannheim (\cite{rm}), where the
binary separation $a$ is obtained by equating the gas dynamical
time-scale (Begelman, Blandford and Rees \cite{bbr}) with the
time-scale for gravitational radiation (Peters and Mathews
\cite{pm}). Here, as previously stated, we release this
assumption and consider $a$ as a free fit parameter.

\begin{figure}[htbp]
\begin{center}
\vspace{11.7cm} \includegraphics{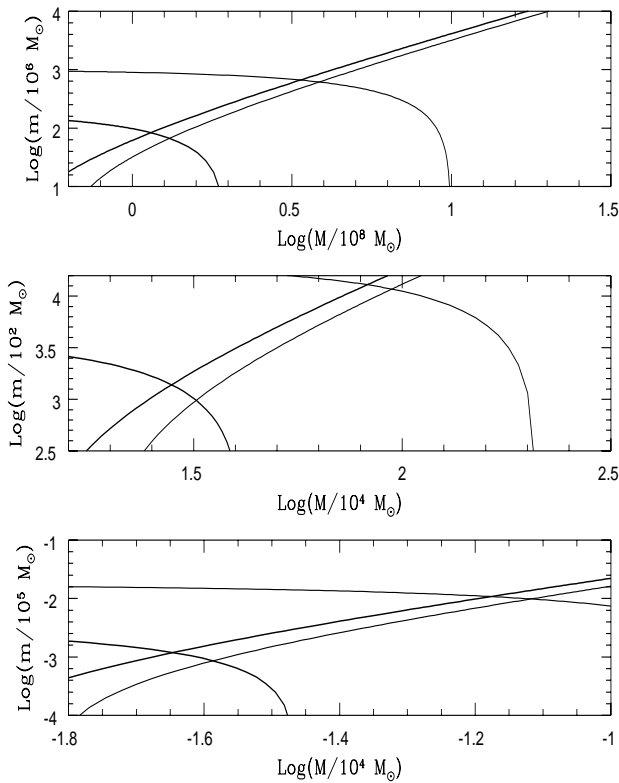} \caption{From the upper to the
bottom panel, the required mass dependence for binary black hole
model is shown in the case of MKN 501, MKN 421 and MKN 766,
respectively. The thick and thin lines represent the conditions
expressed in equations (\ref{massratio1}) and (\ref{massratio2})
with Lorentz factor $\gamma _b= 10$ and $\gamma _b= 15$. The
binary separation $a$ is set to $5\times10^{16}$ cm,
$2\times10^{14}$ cm and $1\times10^{13}$ cm for MKN501, MKN421
and MKN766, respectively. We have set the orbit eccentricity value
equal to $0.5$. The intersection between lines corresponding to
the same Lorentz factor gives the masses of the black holes in
the binary system considered.} \label{mcontromtotale}
\end{center}
\end{figure}

For the three MKN objects considered, the secondary mass $m$ as a
function of the primary one $M$ is shown in Fig.
\ref{mcontromtotale} for selected values of $\gamma_b = 10,15$
and $e=0.5$. For each binary system, the intersection between
lines with the same Lorentz factor determines the masses of the
two black hole components.

\begin{figure}[htbp]
\begin{center}
\vspace{9.cm} \includegraphics{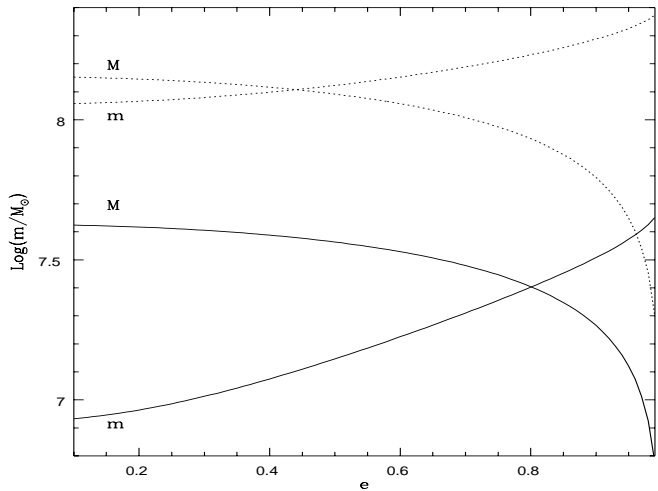} \caption{The dependence of the
primary and secondary mass on the eccentricity of the orbit is
shown for MKN 501 Lorentz $\gamma_b$ factors $\gamma_b=10$
(dotted lines) and $\gamma_b=30$ (solid lines), respectively. Here
we have set $a=8\times10^{16}$ cm. Similar plots can be obtained
for the MKN421 and MKN766.}

\label{mcontrogamma}
\end{center}
\end{figure}

In Fig. \ref{mcontrogamma} (for MKN 501) we show the primary and
secondary black hole masses as a function of the orbit
eccentricity for two values of the Lorentz factor $\gamma_b= 10,
30$ and $a = 8\times 10^{16}$ cm. In this case, the black hole
masses are in the range $10^6-10^9~M_{\odot}$.

\section{Gravitational waves from massive black hole binaries
in Markarian objects and comparison with LISA sensitivity}

In this Section, the massive black hole binaries in the MKN
objects considered in Section 2 are examined as possible sources
of gravitational radiation. Due to the orbital period of the
binary systems considered, it is expected that they emit most of
the gravitational radiation in the low-frequency band ($\nu \ut<
10^{-4}$ Hz). Gravitational waves in this band are only
accessible to experiments based on the Doppler tracking technique
of inter-planetary spacecrafts. The Laser Interferometer Space
Antenna (LISA), consisting of a constellation of three drag-free
spacecrafts at the vertices of an ideal equilater triangle with
sides of $\sim 5\times 10^{6}$ km (Hiscock, Larson, Routzahn and
Kulick \cite{hlrk}), has an optimal sensitivity in the frequency
range covering the band $10^{-5}$ Hz - $3\times 10^{-2}$ Hz
(Vecchio \cite{vecchio}). LISA will be able to carry out a deep
and extensive census of black hole populations in the Universe,
providing an accurate demography of these objects and their
interactions. In particular we expect that binaries of massive
black holes could be detectable by LISA even if the emitted
gravitational waves are characterized by extremely low
frequencies ($\nu \ut< 10^{-4}$ Hz). This could  be the case for
the binaries at the center of the three MKN objects considered
under particular conditions (e.g. for the same values of $a,~e$
and $\gamma_{b}$).

Indeed we expect that in a merging process between galaxies, the
two colliding black holes at their centers would eventually move
on orbits with non-negligible eccentricity. This fact should have
great importance as far as the detection of the emitted
gravitational waves is concerned. In fact, a binary system with
an eccentric orbit emits significantly more power in
gravitational waves than a circular system with the same
semimajor axis. Moreover, eccentric binaries will also emit
gravitational waves in a multitude of harmonics of the orbital
frequency $\omega _k$ associated with the semimajor axis $a$.
Thus, an eccentric system emits gravitational waves at the
frequencies $\omega_n =n \omega_k$ (with $n=1,2,3...$), whereas
circular binaries emit purely in the $n=2$ mode. Peters and
Mathews (\cite{pm}), found that the orbit averaged power emitted
in gravitational waves by a binary system with eccentricity $e$
in the {\it n}th harmonic of the orbital frequency is
\begin{equation}
\frac{dE
(n,e)}{dt}=\frac{32}{5}\frac{G^4M^2m^2(M+m)}{c^5a^5}g(n,e)~,
\label{GWpower}
\end{equation}
where the function $g(n,e)$ depends on $n$ and $e$ by the
following relation
\begin{equation}
\begin{array}{lll}
g(n,e) =\displaystyle{\frac{n^4}{32}}\left\{\left[J_{n-2}(ne)-2eJ_{n-1}(ne)
\right.\right.\\ \\
~~~~~~~~~~~\left.\left.
+\displaystyle{\frac{2}{n}}J_n(ne)+2eJ_{n+1}(ne)-J_{n+2}(ne)\right]^2\right. \\
\\
~~~~~~~~~~~\left.+(1-e^2)\left[J_{n-2}(ne)-2J_n(ne)+J_{n+2}(ne)\right]^2\right.\\
\\
~~~~~~~~~~~\left. +\displaystyle{\frac{4}{3n^2}}
\left[J_n(ne)\right]^2 \right\}~.
\end{array}
\label{gfactor}
\end{equation}
Here $J_n$ (with $n$ integer) is the usual Bessel function of
order $n^{th}$.
 The characteristic spectral amplitude at the frequency
$\omega _n$ can be evaluated as
\begin{equation}
h(\omega_n) \simeq \frac{2}{\pi \omega
_nD}\sqrt{\frac{G}{c^3}\frac{dE (n,e)}{dt}}~,
\label{signalamplitude}
\end{equation}
where $D$ is the distance of the gravitational wave source from
Earth.

\begin{figure*}[htbp]
\vspace{15cm} \includegraphics{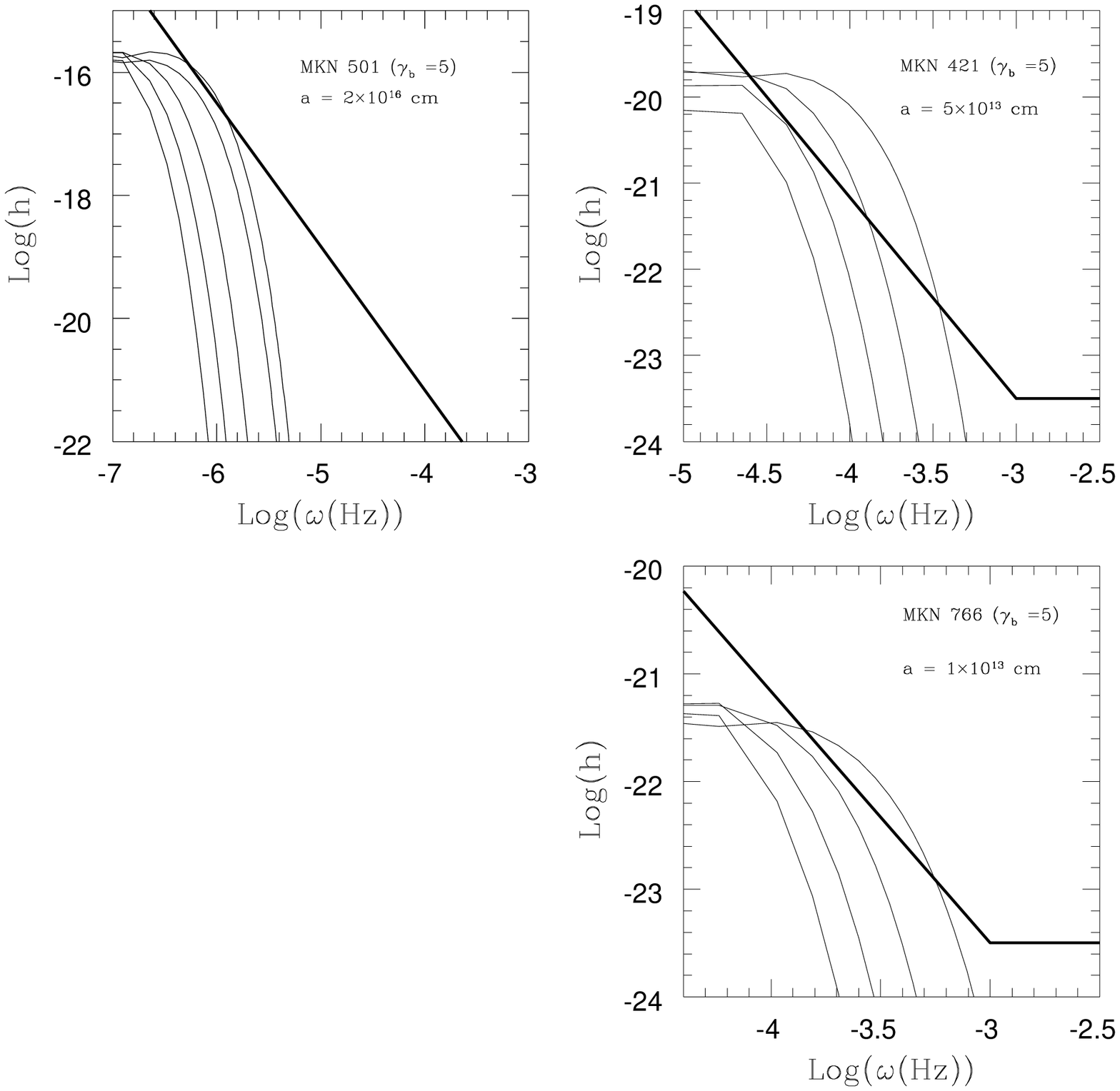}
\caption{The amplitude of the gravitational waves emitted by the
binary black holes fitting all the observational data is reported
as a function of the frequency for the three MKN objects
considered. The continuous lines from left to right in each panel
correspond to orbital eccentricity $e=0.5,~0.6,~0.7,~0.8$,
respectively (for MKN 501 we added a curve corresponding to
$e=0.85$). The chosen orbital radius $a$ is indicated in each
panel. The solid oblique line represents the LISA sensitivity
with an integration time of 5 years.}
\label{hvsftot}
\end{figure*}

Equations (\ref{GWpower})-(\ref{signalamplitude}) allow us to
evaluate the expected spectrum of the gravitational waves emitted
by the massive black hole binaries assumed to be in the center of
the considered MKN objects. Assuming $\gamma _b=5$, Fig.
\ref{hvsftot} shows the expected gravitational wave amplitude $h$
as a function of the frequency $\omega$, for different values of
the binary orbit eccentricity. In each panel of the Figure we give
for comparison the LISA sensitivity curve (the oblique solid line)
for an integration time of $5$ years. As can be noted, in several
cases the expected gravitational waves emitted by the considered
binary systems may be detected by LISA satellite. Eccentricity
values up to $0.8$ are not necessarily too extreme. In fact, if a
black hole binary forms in a merging process, a very high value
of eccentricity is expected. Of course, due to the gravitational
wave emission the orbit tends to circularize but this should
happen within a time-scale of the same order of the magnitude of
the merging time-scale (Peters \cite{peters}).

\begin{figure*}[htbp]
\vspace{9.cm} \includegraphics{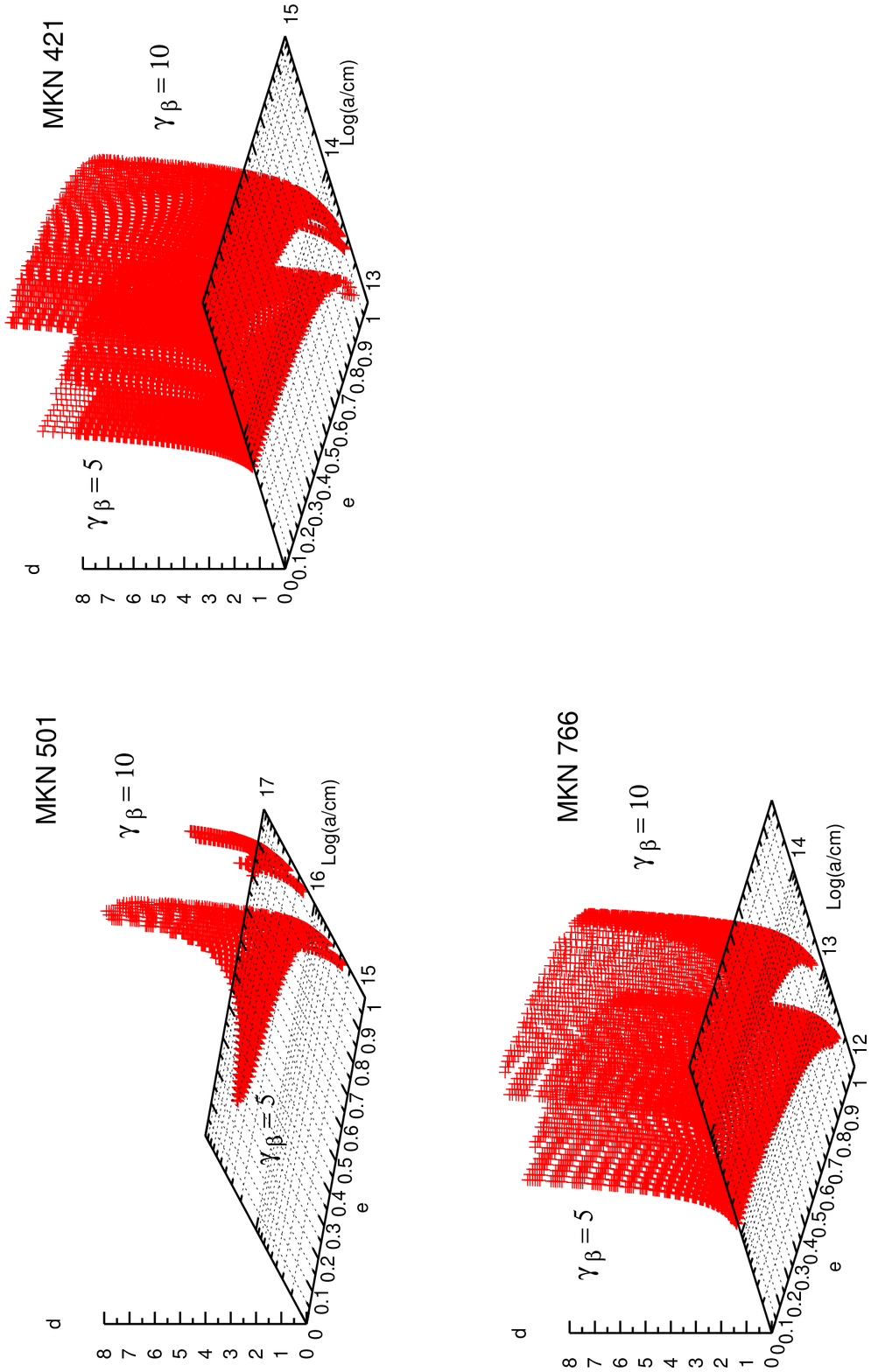} \caption{In the 3D scatter
plot, the binary semimajor axis $a$ is shown as a function of the
orbital eccentricity $e$ for different values of Lorentz factor
$\gamma_{b}$, for MKN 501, MKN 421 and MKN 766, respectively. Each
point in the evidenced regions of the a-e plane fits the
observations and satisfies eqs. (\ref{massratio1}) and
(\ref{massratio2}). On the vertical axis, the detection parameter
$d$ is also shown (for detalis see text).} \label{modelli2}
\end{figure*}

In Fig. \ref{modelli2}, the binary semimajor axis $a$ is shown as
a function of the orbital eccentricity $e$ and for two different
Lorentz factors $\gamma_{b}$ for MKN 501, MKN 421 and MKN 766,
respectively. Each point in the 3D scatter plot fits the
observations and satisfies eqs. (\ref{massratio1}) and
(\ref{massratio2}). Each point on the plot corresponds to a
different distance $d=h(\omega_n)/h_{LISA}(\omega_n)$ (vertical
axis)  between the gravitational wave emitted spectrum and the
LISA sensitivity curve for an integration time of 5 years (the
minimum distance between the curves diminishes as $d$ increases).
In Fig. \ref{modelli2}, only models with $\inf{d(\omega_n)}\geq
0.1$ are shown. Models corresponding to $d\geq 1$ in the plot
have to be considered as observable by the LISA satellite. We
notice that models corresponding to higher values of the
parameter $d$ have typically higher masses and lower coalescing
times with respect to those with lower $d$ values.  In any case,
black hole masses are below $10^9~M_{\odot}$.

\section{Discussion}

We have considered some astrophysical implications of the
existence of massive black hole binary systems at the center of
MKN objects. These systems are expected to be fairly common in the
Universe as a consequence of merging processes between galaxies.
In fact, following the evolutionary scenario, giant elliptical
galaxies are thought to be the result of merging between spirals.
Since almost each galaxy contains a massive black hole in its
center, merging process would naturally lead to the formation of
massive black hole binaries.

Currently, the only method to probe the existence of massive
binary black holes is to search for double nuclei in the galactic
centers. However, as pointed out by Rieger and Mannheim
(\cite{rm}) and Yu (\cite{yu}), a periodic behaviour in the
observed radio, optical, $X$-ray or $\gamma$-ray light curves
could be considered as the signature of a massive binary black
hole.

In this paper, we focused on the possibility of explaining the
intensity variations observed in some MKN objects as the
consequence of the motion of binary black holes around the common
center of mass. We have generalized the treatment of Rieger and
Mannheim (\cite{rm}) considering in detail the effects of the
orbital eccentricity.

At least three MKN objects (MKN 501, MKN 421 and MKN 766) reveal a
possible periodic behaviour in the observed signal. In
particular, it has been recently addressed whether MKN 501 hosts a
massive black hole binary system (Rieger and Mannheim \cite{rm}).
The same likely holds for MKN 421 and MKN 766. The model
described in Section 2 allows us to estimate the mass of each
black hole in the binary system (see Fig. \ref{mcontromtotale})
as a function of the Lorentz factor $\gamma_b$, of the orbital
radius $a$ and eccentricity $e$.

If massive binary black holes really exist in the centers of the
MKN objects considered above, they should also emit low frequency
gravitational waves typically with $\omega \ut< 10^{-4}$ Hz. In
Fig. \ref{hvsftot} we have presented the gravitational wave
spectrum emitted by binaries that fit all the observational data
(flux ratio $f$, the signal periodicity $P_{obs}$ and the source
spectral index $\alpha$). As one can note, for some values of $a$
and $e$ one can obtain curves above the sensitivity threshold of
the LISA detector (we assume an integration time of $5$ years).

It is obvious that binary systems emitting gravitational waves
detectable by the LISA satellite have short coalescing times. For
example, in the case of MKN 501 the average coalescing time for
the models represented in Fig. \ref{modelli2} (for $\gamma_b=5$)
is $\sim 10^3$ years and tends to decrease for higher values of
the parameter $d$.

However, for signal detection, some caution is needed since the
LISA sensitivity curve we use in Figs. \ref{hvsftot} and
\ref{modelli2} does not take into account the noise from the
galactic binaries containing two compact objects. Especially
important are galactic white dwarf binaries (Hils and Bender
\cite{hb}, Nelemans, Yungelson and Portegies-Zwart \cite{nyp},
Nelemans, Portegies-Zwart and Verbunt \cite{npv}), primordial
black holes (Hiscock \cite{hiscock}) and white dwarf binaries
(Hiscock, Larson, Routzahn and Kulick \cite{hlrk}) in the
galactic halo. These binaries would give rise to a confusion
noise which lies well above LISA's sensitivity curve in the
frequency range of interest between $10^{-4}$ Hz and $10^{-3}$ Hz
(see e.g. Fig. 3 in Nelemans, Yungelson and Portegies-Zwart
\cite{nyp}). This noise would make even more difficult the
detection of the considered massive black hole binary systems so
that, in order to have detectable signals, one would need more
extreme conditions (e.g. higher values of the parameter $d$ or,
equivalently, shorter coalescing times).

As a final remark we notice that it is possible that there are a
lot of hard binary black holes at the center of galaxies, MKN
objects, Seyfert galaxies, etc., and that we can observe directly
(from the periodicity of lightcurves at high energy) only a small
fraction of them. If this is so, detection of gravitational waves
from such systems would be a means to detect them. It is worth
noting that the number of intervening galaxies between Earth and
a galaxy at redshift $z$ is
\begin{equation}
N\simeq 6\times10^{-3}\left(\frac{R}{3~{\rm kpc}}\right)^2
\left(\frac{N_g}{0.1~{{\rm Mpc}^{-3}}}\right)[(1+z)^{3/2}-1]~,
\end{equation}
where $R$ is the typical galaxy radius and $N_g$ the average
number density of galaxies. Therefore, it is expected that at
least in some cases gravitational lenses, displaced along the
line of sight of a gravitational wave source, should distort the
path of the gravitational waves just as they distort the path of
the observed light. For gravitational waves with frequency $\nu
\ut < 10^{-4}$ Hz, the LISA angular resolution is below $10^{-2}$
steradians (Cutler \cite{cutler}). Consequently, the lensed
source is always seen as an amplitude amplification (i.e. as a
microlensing event) and the observed amplitude, at the frequency
$\omega _n$, is (De Paolis, Ingrosso and Nucita \cite{din})
\begin{equation}
h^{max}(\omega _n) \simeq \sqrt{A^{max}} h(\omega _n)
\end{equation}
where $A^{max}$ is the amplification parameter depending on the
geometry of the microlensing configuration. This amplification
could be in some cases of the order of $10$ or more, thereby
increasing the number of detectable binary black holes at
galactic centers by future gravitational wave detectors like LISA.
Even more interesting would be the effect of the gravitational
wave time-delay due to the different paths traveled by the
deflected signals. This time-delay, of about one year as for the
lensed Quasars, would give rise to a frequency difference due to
the binary evolution with time. This effect could produce a
``beating pattern'' in the wave amplitude which could be searched
for in future gravitational wave experiments.

\acknowledgements{We would like to thank Prof. Asghar Qadir for
interesting discussions about the problem. We also warmly
acknowledge an anonymous referee for pointing out some important
issues.}


\begin{thebibliography}{99}
\bibitem[1999]{aab}
F. Aharonian, A.G. Akhperjanian, J.A. Barrio et al., 1999, A\&A
349, 29
\bibitem[1980]{bbr}
M.C. Begelman, R.D. Blandford and M.J. Rees, 1980, Nat 287, 307
\bibitem[2000]{bkt}
T. Boller, R. Keil, J. Tr$\ddot{u}$mper et al., 2001, A\&A 365,
134
\bibitem[2000]{bsg} W.
Brinkmann, S. Sembay, R.G. Griffiths et al., 2001, A\&A 365, 162
\bibitem[1997]{cbb}
M. Catanese, S.M: Bradbury, A.C. Beslin et al., 1997, ApJ 487,
L143
\bibitem[1995]{cw}
J.E. Conway and J.M. Wroble, 1995, ApJ 439, 98
\bibitem[1998]{cutler}
C. Cutler, 1998, Phys.Rev. D 57, 7089
\bibitem[2001]{din}
F. De Paolis, G. Ingrosso and A. A. Nucita, 2001, A\&A, 366, 1065
\bibitem[1987]{fitchett}
M. Fitchett, 1987, MNRAS, 224, 567
\bibitem[1988]{gwb}
I.M. George, R.S. Warwick and G.E. Bromage, 1988, MNRAS, 232, 793
\bibitem[1999]{gvm}
M. Guainazzi, G. Vacanti, A. Malizia et al., 1999, A\&A 342, 124
\bibitem[1998]{haehnelt}
M.G. Haehnelt, 1998, {\it Laser Interferometer Space Antenna,
Second Int. LISA Symp. on the Detection and Observation of
Gravitational Waves in Space}, Ed. W.M. Folkner, AIP Conf.Proc.
456, p. 45
\bibitem[1998]{hhi}
N. Hayashida, H. Hirasawa, F. Ishikawa et al., 1998, ApJ 504, L71
\bibitem[1998]{hiscock}
W.A. Hiscock, 1998, ApJ 509, L101
\bibitem[2000]{hlrk}
W.A. Hiscock, S.L. Larson, J.R. Routzahn and B. Kulick, 2000, ApJ
540, L5
\bibitem[1997]{katz}
J.I. Katz, 1997, ApJ 478, 527
\bibitem[1995]{kr}
J. Kormendy and D. Richstone, 1995, ARA\&A, 33, 581
\bibitem[1999]{kdk}
D. Kranich, O.C. deJager, M. Kestel et al., 1999, in: {\it Proc.
of 26th International Cosmic Ray Conference (Salt Lake City)} 3,
p.358
\bibitem[2000]{hb} D. Hils and P.L. Bender, 2000, ApJ 537, 334
\bibitem[1996]{lv}
H.L. Letho and M.J. Valtonen, 1996, ApJ 460, 207
\bibitem[1999]{mft}
L. Maraschi, G. Fossati, F. Tavecchio et al., 1999,
Astropart.Phys. 11, 189
\bibitem[1993]{mms}
S. Molendi, T. Maccaro and S. Schaeidt, 1993, A\&A 271, 18
\bibitem[1999]{npv}
G. Nelemans, S. F. Portegies-Zwart and F. Verbunt, 1999, preprint
astro-ph/9903255
\bibitem[2001]{nyp}
G. Nelemans, L. R. Yungelson and S. F. Portegies-Zwart, 2001,
A\&A 375, 890
\bibitem[1999]{nhc}
D. Nishikawa, S. Hayashi, N. Chamoto et al., 1999, in: {\it Proc.
of 26th International Cosmic Ray Conference (Salt Lake City)} 3,
p.354
\bibitem[1964]{peters}
P.C. Peters, 1964, Phys. Rev. 136, 1224
\bibitem[1963]{pm}
P.C. Peters and J. Mathews, 1963, Phys. Rev. 131, 435
\bibitem[1998]{pbf}
R.J. Protheroe, C.L. Bhat, P. Fleury et al., 1998, in:{\it Proc.
25th International Cosmic Ray Conference (Durban)} 8, p 317
\bibitem[1992]{pac}
M. Punch, C.W. Akerlof, M.F. Cawley et al., 1998, Nature 358, 477
\bibitem[1984]{rees}
M. Rees, 1984, ARA\&A,  22, 471
\bibitem[1998]{rab}
D. Richstone, E.A. Ajhar and R. Bender, 1998, Nature 395, 14
\bibitem[2000]{rm}
F.M. Rieger and K. Mannheim, 2000, A\&A 359, 948
\bibitem[1983]{st}
S.L. Shapiro and S.A. Teukolsky, 1983, {\it Black Holes, White
Dwarfs, and Neutron Stars}, John Wiley and \& Sons, New York
\bibitem[1977]{smart}
W.M.Smart, 1977, {\it Textbook on Spherical Astronomy}, Cambridge
University Press, Cambridge
\bibitem[1999]{spada}
M. Spada, 1999, Astropart. Phys. 11, 59
\bibitem[1988]{shv}
A. Sillinpa$\ddot{a}\ddot{a}$, S. Haarala, M.J. Valtonen et al.,
1988, ApJ 325, 628
\bibitem[1995]{up}
L.M. Urry and P. Padovani, 1995, PASP 107, 803
\bibitem[1999]{vecchio}
A. Vecchio, 1999, in: {\it Gravitational Waves, Third E. Amaldi
Conference}, AIP Conference Proceedings, vol. 523 p. 238, AIP
Press
\bibitem[1999]{vr}
M. Villata and C.M. Raiteri, 1999, ApJ 347, 30
\bibitem[1997]{white}
S.D.M. White, in: B$\ddot{o}$rner G., Gottlober S. (eds), {\it The
evolution of the Universe: Report of the Dahlem Workshop} (Berlin,
1997), p. 227
\bibitem[2001]{yu}
Q. Yu, 2001, pre-print astro-ph/0109530
\bibitem[2001]{yulu}
Q. Yu and Y. Lu, 2001, A\&A 377, 17
\end{thebibliography}
\end{document}